\documentclass[12pt,preprint,iop]{emulateapj}
\usepackage{epsfig, natbib} 
\usepackage{mathrsfs,amssymb}
\usepackage{graphicx,latexsym,footnote}
\newcommand{\am}{\mbox{$'$}}

\newcommand{\as}{\mbox{$''$}}

\newcommand{\degree}{\mbox{$^{\circ}$}}
\newcommand{\etal}{{\it et~al.\/}}
\newcommand{\fesc}{\mbox{$f_{\rm esc}$}}

\newcommand{\Hline}[1]{\mbox{H{\footnotesize {#1}}}}
\newcommand{\Halpha}{\Hline{\mbox{$\alpha$}}}

\newcommand{\kms}{\mbox{km\thinspace s$^{-1}$}}

\newcommand{\mum}{\mbox{$\mu \rm m$}}

\newcommand{\SthStw}{\mbox{[S~III]/[S~II]}}

\slugcomment{*** Accepted to ApJ Letters 28 Sept 2011***}

\shorttitle{An Ionization Cone in NGC 5253}
\shortauthors{Zastrow \etal}

\begin{document}

\title{An Ionization Cone in the Dwarf Starburst Galaxy NGC 5253\footnotemark[1]}
\footnotetext[1]{This paper includes data gathered with the 6.5 meter Magellan Telescopes located at Las Campanas Observatory, Chile.}

\author{Jordan\ Zastrow\altaffilmark{2},
        M.S.\ Oey\altaffilmark{2},
        Sylvain\ Veilleux\altaffilmark{3},
        Michael\ McDonald\altaffilmark{4},
        Crystal\ L.\ Martin\altaffilmark{5}}

\altaffiltext{2}{Department of Astronomy, University of Michigan, 830 Dennison Building, 500 Church Street, Ann Arbor, MI, 48109-1042, USA; JZ: jazast@umich.edu}
\altaffiltext{3}{Department of Astronomy, University of Maryland, College Park, MD 20742, USA}
\altaffiltext{4}{Kavli Institute for Astrophysics and Space Research, MIT, Cambridge, MA 02139, USA}
\altaffiltext{5}{Department of Physics, University of California, Santa Barbara, CA 93106, USA}


\begin{abstract}

There are few observational constraints on how the escape of ionizing 
photons from starburst galaxies depends on galactic parameters.  Here, 
we report on the first major detection of an ionization cone 
in NGC 5253, a nearby starburst galaxy.  This 
high-excitation feature is identified by mapping the emission-line 
ratios in the galaxy using [S III] $\lambda$9069, [S II] $\lambda$6716, 
and \Halpha\ narrow-band images from the Maryland-Magellan 
Tunable Filter at Las Campanas Observatory.  The ionization cone 
appears optically thin, which is suggestive of the escape of 
ionizing photons.  The cone morphology is 
narrow with an estimated solid angle covering just 3\% of $4\pi$ 
steradians, and the young, massive clusters of the nuclear starburst can
easily generate the radiation required to ionize the cone.
Although less likely, we cannot rule out the possibility of an obscured 
AGN source.  An echelle spectrum along the minor axis shows complex
kinematics that are consistent with outflow activity.
The narrow morphology of the ionization cone supports the scenario 
that an orientation bias contributes to the difficulty in detecting 
Lyman continuum emission from starbursts and Lyman break galaxies.
  
\end{abstract}

\section{Introduction} \label{s:intro}

The fate of ionizing photons in starburst galaxies is a critical 
issue for our understanding of cosmic reionization.  
Early-epoch star-forming galaxies are the 
most likely candidate source for the required ionizing radiation 
\citep[e.g.,][]{b:madau_apj99,b:fan_aj01}.  However, 
quantifying the escape fraction of ionizing radiation (\fesc) 
has proven to be challenging.  

Generally, the interstellar medium (ISM) is optically thick to 
Lyman continuum.  As discussed by \citet{b:heckman_apj11}, 
the average hydrogen column density in galaxies ranges from 
a factor $10^4$\ to $10^7$ times higher than the column that 
produces unity optical depth in the Lyman continuum.  
Thus, \fesc\ will be dependent on the morphology of the 
ISM.  For example, models have shown that a clumpy ISM yields 
a higher \fesc\ than a smoothly varying 
medium \citep[e.g.,][]{b:ciardi_mnras02, b:fernandez_apj11}.  
For ionizing radiation to escape, low density paths out 
of the galaxy must be created.  In starburst galaxies, the mechanical 
feedback from massive stars creates low density bubbles in the 
ISM, which can then go on to break out of the galactic disk, facilitating 
the escape of radiation \citep[e.g.,][]{b:mac-low_apj88,b:clarke_mnras02}.
However, the influence of feedback on \fesc\ is not 
straightforward.  Simulations show that the 
very shells and bubbles that generate low-density holes in the 
ISM can initially trap ionizing photons before break-out, which will affect 
\fesc\  \citep{b:fujita_apj03, b:dove_apj00}. 

Understanding this problem is complicated 
by the small number of galaxies detected with significant \fesc, 
despite large effort by the community to measure it both 
locally and at high redshift.  Direct detection of excess Lyman 
continuum has been found in a subset of $z \sim 3$ Lyman-break 
galaxies \citep[e.g.][]{b:steidel_apj01}. However, the fraction of 
galaxies with high \fesc\ is low, on the order of 
10\% \citep[e.g.,][]{b:iwata_apj09,b:shapley_apj06}.  
For local starbursts, the detection rate drops, and most 
studies obtain upper limits to \fesc\ of a few percent 
\citep[e.g.,][]{b:leitherer_apjl95,b:heckman_apj01,b:grimes_apjs09,b:siana_apj10}.  

One possible explanation for the low detection rate is an orientation bias.  
In the paradigm outlined above,  feedback processes will create 
passageways perpendicular to the galaxy plane.  It follows that 
ionizing photons will preferentially escape in that direction with an 
opening angle dependent on galactic conditions \citep{b:dove_apj00,b:gnedin_apj08,b:fernandez_apj11}.  
Direct measurements of \fesc\ will thus depend on the orientation 
of the galaxy to our line-of-sight.  This model is conducive to the
formation of photoionized ionization cones.  While photoionized 
emission has been observed in outflows near starburst 
nuclei \citep[e.g.,][]{b:sharp_apj10,b:shopbell_apj98}, clear evidence for 
ionization cones extending well into the galaxy halo has not yet been observed.  Here,
we report a new detection of such an ionization cone in a starburst
galaxy, NGC 5253.  Its narrow opening angle suggests that orientation 
may play an important role in the detectability of escaping Lyman continuum radiation.
\\
\section{Observations}\label{s:obs}
   
NGC~5253 is a nearby, dwarf galaxy undergoing intense, centrally 
concentrated star formation.  The most recent episode of star formation 
produced massive super star clusters with ages of just a few 
Myr \citep[e.g.,][]{b:calzetti_aj97}.  At a distance of 3.8 Mpc \citep[][]{b:sakai_apj04}, 
we are able resolve structure down to tens of parsecs.  NGC~5253 is
well studied across many wavelengths, providing much ancillary 
data to augment this study.   

To identify channels where Lyman continuum radiation might 
escape from the galaxy, we apply the technique of ionization 
parameter mapping \citep[e.g., Pellegrini et al., in preparation;][]{b:pogge_apj88}.  
The ionization parameter is a measure of the ionizing
radiation energy density relative to the gas density, and it can be mapped
by emission-line ratios that probe high- vs low-ionization species.
Regions that are optically thick to ionizing 
radiation show a low-ionization boundary between the highly ionized
regions close to the ionizing source and the neutral environment;
whereas optically thin regions maintain high excitation throughout.
For this study, we use [S III] $\lambda 9069$ and  [S II] $\lambda
6716$ to map the high- and low-ionization species, respectively.

We obtained narrow-band images of NGC~5253 on the nights 
of 2009 July 9--11 using the Maryland-Magellan 
Tunable Filter \citep[MMTF;][]{b:veilleux_aj10}.  The 
bandpasses include \Halpha\ and narrow-band continuum imaging
observed at the redshift of NGC~5253.  They are given
in Table \ref{t:galparams}, together with
their exposure times.  MMTF uses a 
low-order Fabry-Perot etalon and is mounted on the imaging 
spectrograph, IMACS, of the Magellan Baade telescope at 
Las Campanas Observatory.  The IMACS f/2 camera has an 
8K $\times$ 8K Mosaic CCD with a pixel scale of 0.2\as\ per pixel.  
For these observations, MMTF was used at low etalon spacings to 
provide a monochromatic field of view of 11.5\am.  We additionally 
obtained a long-slit spectrum of NGC 5253.  The seeing conditions 
ranged from 0.5\as--0.7\as\ for the [S II] and [S III] images and 
1.5\as--2\as\ for \Halpha.  

The bias subtraction, flat-fielding, and sky subtraction 
are accomplished using version 1.4 of the MMTF data reduction 
pipeline,\footnote{http://www.astro.umd.edu/\textasciitilde 
veilleux/mmtf/datared.html} described in \citet{b:veilleux_aj10}.   
The \Halpha-continuum image is first flux calibrated using photometry 
from standard star observations.  Then, the long slit spectrum is 
corrected for instrument response using standard star 
spectra, and flux calibrated using the flux-calibrated, 
\Halpha-continuum image.  Finally, the other emission-line images 
are normalized to the emission measured across each bandpass in the 
flux-calibrated, long slit spectrum.  Systematic errors for the flux 
calibration are about a factor of two.  The images are registered and 
rescaled using standard IRAF\footnote{IRAF is distributed by NOAO, 
which is operated by AURA, Inc., under cooperative agreement 
with the National Science Foundation.} tasks, then continuum 
subtracted.  Galactic extinction is corrected assuming 
$E(B-V) = 0.0475$ across the field \citep{b:burstein_aj82} and 
the extinction law from \citet{b:cardelli_apj89}.  The internal 
extinction is known to be variable throughout the galaxy
\citep{b:calzetti_aj97,b:caldwell_apj89,b:cresci_aap05}, and
its effects are discussed in \S\ 3.1.

\begin{deluxetable*}{ccccc}
\tablewidth{0pt}
\tabletypesize{\footnotesize}
\tablecaption{Observations \label{t:galparams}}
\tablehead{\colhead{Emission} & \colhead{Wavelength } &\colhead{Effective} &  \colhead {Exp. Time} &\colhead{Flux\tablenotemark{a}} \\
\colhead{line}& \colhead{(\AA)} & \colhead{Bandpass (\AA)}& \colhead{}& \colhead{ (10$^{-13}$\ erg s$^{-1}$\ cm$^{-2}$ ) }}
\startdata
[S~III] & 9081 & 27.0 & $5 \times 1200$ s & $0.39 $\\
continuum & 9180 & 27.0 & $5 \times 1200$ s & ... \\
$\rm [S~II]$ & 6726 & 16.2 &$5 \times 1200$ s & $ 0.59 $ \\
continuum & 6680 & 16.2 & $5 \times 1200$ s & ... \\
\Halpha& 6572 &18.5 & $3 \times 1200$ s&$12.9 $\\
continuum & 6680 & 18.5 & $3 \times 1200$ s & ...
\enddata
\tablenotetext{a}{\scriptsize Continuum-subtracted flux measured in the ionization cone using the DS9 funtools module.}
\end{deluxetable*}

\begin{figure}[h]
\includegraphics[width=8.5cm]{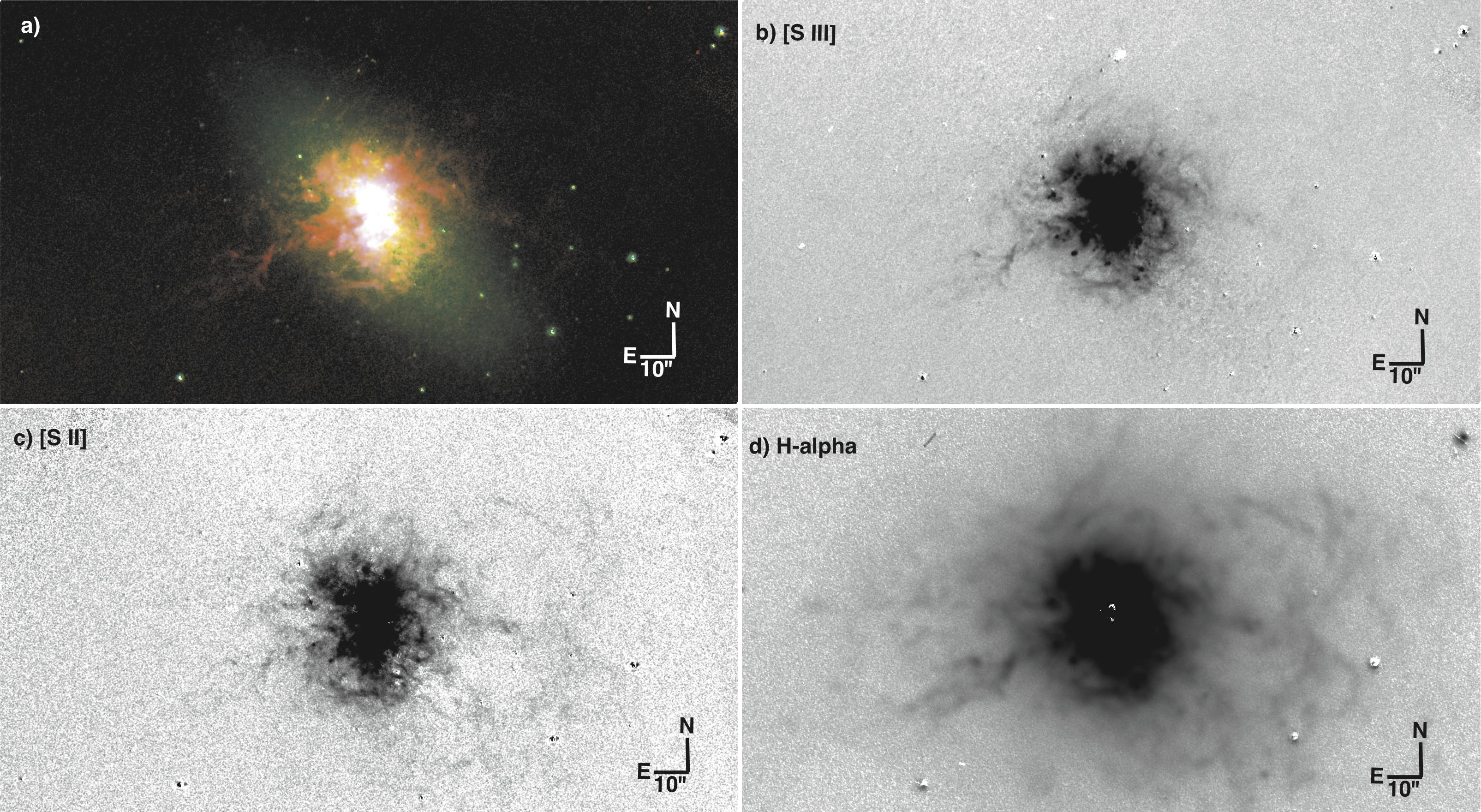}
\caption{\emph{a)}\ Composite image of NGC~5253. Red, blue and green 
correspond to [S~III], [S~II], and continuum at $\lambda6680$, 
respectively. Ionization cone extends ESE from the nuclear starburst. 
Panels \emph{b, c,} and \emph{d} show individual continuum-subtracted,
[S~III], [S~II], and \Halpha, respectively.
At a distance of 3.8 Mpc, 10\as = 180 pc. \label{f:threecolor}}
\end{figure}

A three-color composite of our data is presented in Figure 
\ref{f:threecolor}, along with the continuum-subtracted, 
emission-line images.  Sky background and noise properties 
of the images are measured using the median and standard 
deviation in the flux of fifteen $20\as\times 20\as$\ boxes 
around the galaxy.  In \Halpha, we have a  3$\sigma$\ 
confidence detection limit of $4.1\times 10^{-18}\rm\ erg\thinspace 
s^{-1}\thinspace cm^{-2}$.  The ionized gas in the 
galaxy has a roughly spherical distribution around the nucleus, 
with networks of loops and filaments particularly noticeable to the south 
and northwest \citep[e.g.,][]{b:marlowe_apj95}.   

\section{Ionization Cone in NGC 5253}\label{s:cone}

\subsection{Morphology and Excitation of the Ionization Cone}

\begin{figure}[h]
\includegraphics[width=8.5cm]{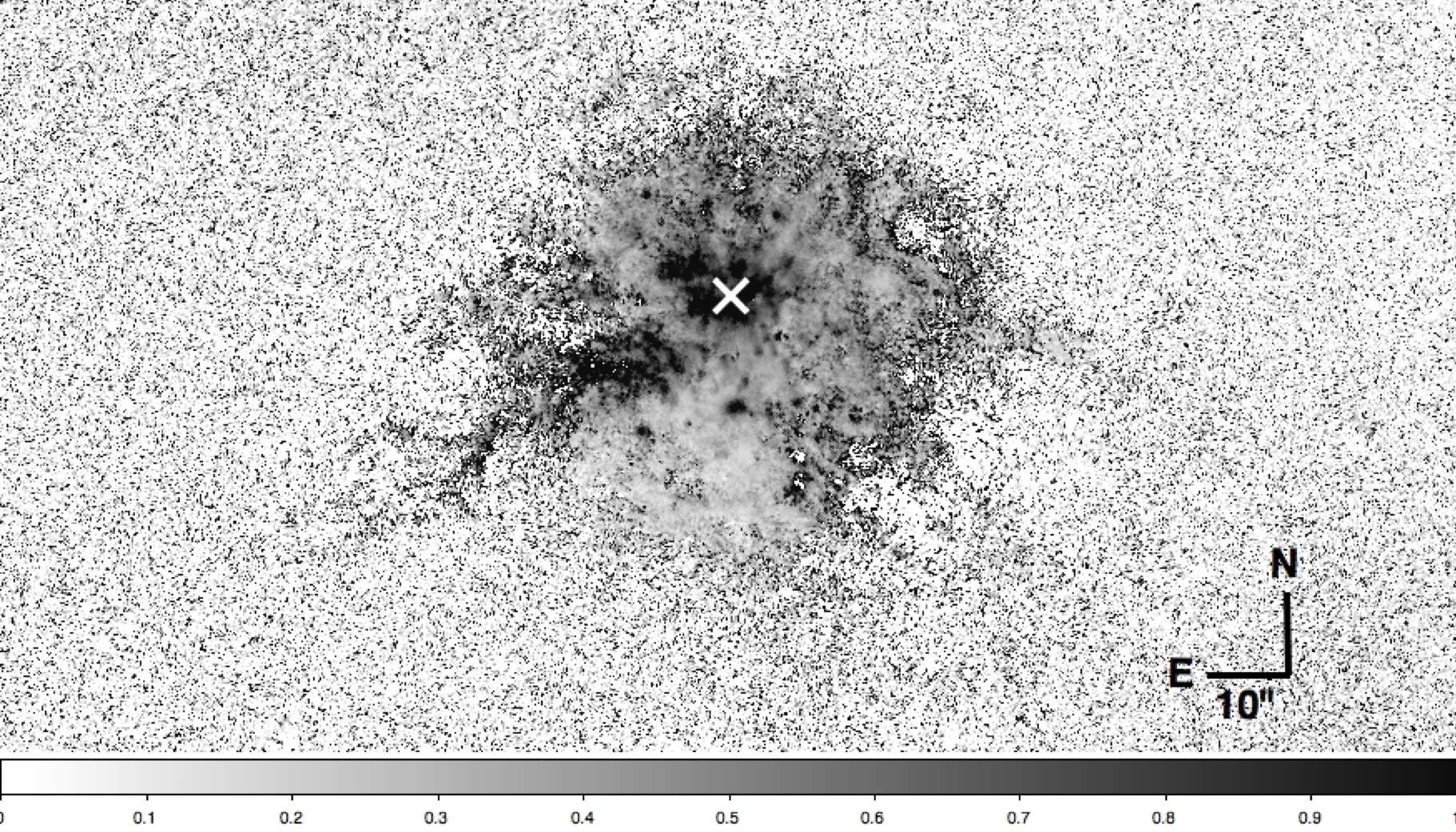}\\
\caption{\SthStw\ ratio map for NGC~5253. 
Dark indicates high values of \SthStw.  
\label{f:ion_param}}
\end{figure}

NGC 5253 shows a stunning detection of an ionization cone in a starburst
galaxy.  In Figure \ref{f:threecolor}, the ionization cone stands out 
beautifully, extending ESE along the minor axis.  In the ratio map, 
Figure \ref{f:ion_param}, the ionization cone is identified by 
its high excitation relative to the rest of the galaxy.  This high 
excitation has also been observed in [O~III]/[S~II] \citep{b:calzetti_aj99}.  
It has an opening angle $< 40$\degree\ with a narrow appearance 
at large radii.  As the distance from the galaxy increases, 
the morphology of the cone becomes filamentary.  While the cone 
appears to be one-sided, we note that extended [S~III] emission 
to the northwest may suggest a bi-cone (Figure~\ref{f:threecolor}).
 
\begin{figure}[h]
\includegraphics[width=8.5cm]{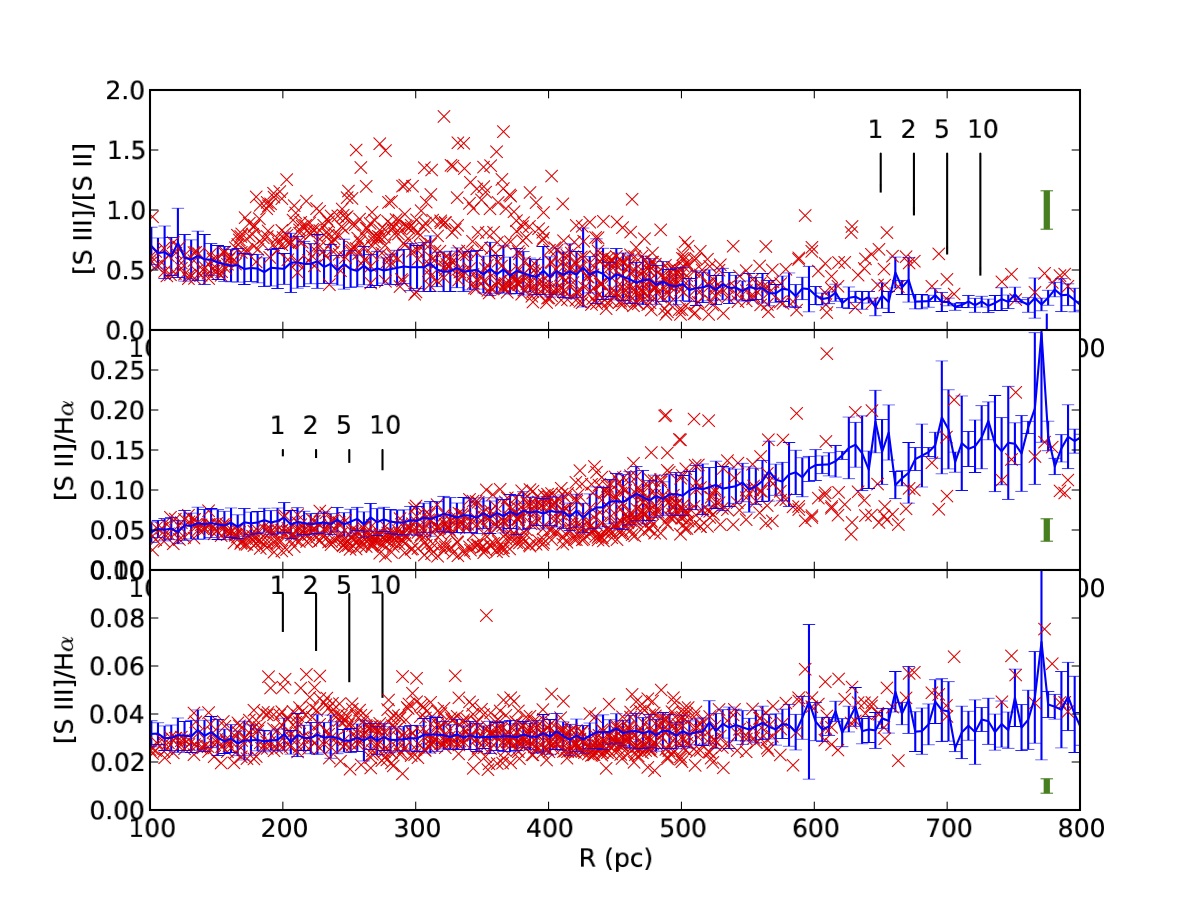}
\caption{Emission-line ratios vs. galactocentric radius.
Red represents ionization cone emission.  Blue line is the mean galactic 
value in radial bins, with bars noting one standard deviation. Representative 
measurement error is shown by the green error bar.  The black lines 
indicate the effect of different levels of extinction ($A_V = 1, 2, 5, 10$). 
Top, middle, and bottom panels show \SthStw, [S~II]/\Halpha,
and [S~III]/\Halpha, respectively. \label{f:Uplot}}
\end{figure}

The changes in ionization parameter as a function of radius 
are shown more quantitatively in Figure \ref{f:Uplot}.  
These plots are generated using $3\times3$ binning of the 
continuum-subtracted images, and a 3$\sigma$\ detection 
cut-off in all three emission lines.  The ionization cone, shown in red, 
is defined as the region between P.A. = 102\degree--140\degree\ using 
$\alpha = \rm 13^h39^m55^s.7, \delta = -31\degree38\am24\as$\ 
(J2000) as the apex.  This point, marked by the cross in 
Figure \ref{f:ion_param}, is the location of a likely source for the 
ionizing radiation, as discussed below.  For comparison, the blue 
line is representative of the rest of the galaxy.   

The top panel of Figure~\ref{f:Uplot} clearly shows a strong excess 
in the \SthStw\ ratio for the ionization cone at lower galactocentric
radii $R$, and a clear excess that continues to $R\sim600 - 700$ pc.
The middle and lower panels show that this excess cannot be explained
solely by the well-known increasing gradient in [S~II]/\Halpha\ that
is found in the diffuse, ionized gas of star-forming galaxies
\citep[e.g.,][]{b:rand_apj98,b:haffner_apj99}.
Furthermore, Figure~\ref{f:threecolor} shows that the primary
star-forming region extends only to $\sim500~$ pc, whereas 
photoionized material associated with the cone clearly extends beyond.
This strongly suggests that there is not enough gas in the outer
halo to absorb all the ionizing photons, and that they therefore
may escape the galaxy.

As noted above, the extinction throughout the starburst is 
variable \citep{b:cresci_aap05}, and
might enhance the observed \SthStw\ ratio relative to the intrinsic one.
In particular, a dust lane with associated CO along 
the minor axis is coincident with the location of the 
cone \citep{b:calzetti_aj97, b:caldwell_apj89,b:meier_aj02}.  
However, Figure 3 shows that we need $A_V\sim 3-10$ mag
to explain the observed \SthStw\ ratios entirely by 
the dust lane, which is much more than its observed extinction 
\citep[$A_V \sim 2.2$\ mag,][]{b:calzetti_aj97}.  If there were that much 
extinction, we would expect a much lower flux along the cone, which
is not seen in the emission-line images in Figure \ref{f:threecolor}.  
Additionally, the [S~II]/\Halpha\ ratio observed in the cone cannot be 
explained by extinction.  We also note that the highest extinction will 
likely be strongest at the smallest $R$.  

Another possible effect is a bias in our observed \SthStw\ ratio 
caused by strong variations in electron density.  
This affects the [S~II]$\lambda6716/\lambda6731$ ratio,
so that our observed $\lambda 6716$ may underestimate the total 
[S~II] at high density.  This error again will be largest at
smaller radii where densities are larger, and leads to difference of
$\sim25\%$ between a density of $10^2$ and $10^3 \rm\ cm^{-3}$.  
In our long-slit spectrum, [S~II] $\lambda\lambda 6716/6731$ ratio 
ranges from 1.4 to 1.0, which correspond to densities of
$\lesssim 100 \rm\ cm^{-3}$ and $\sim 500\ \rm cm^{-3}$, respectively.  
Since this is a smaller variation than between $10^2$ and $10^3 \rm\ cm^{-3}$, 
the error on our observed [S~II] due to density variations is likely much less than 25\%.

The measured emission-line fluxes in the ionization cone 
(Table \ref{t:galparams}) can set limits on the ionizing 
source.  The \Halpha\ flux corresponds to an \Halpha\ luminosity 
$L(\Halpha) = 2.2 (\pm 1.1) \times 10^{39}\ \rm erg\thinspace s^{-1}$.   
This is $\sim 6\%$ the total $L(\Halpha)$ in the galaxy 
\citep[][corrected for distance]{b:marlowe_apj95}, from which
we obtain a lower limit of $\log(Q_0) \geq\ 51.19$ on the rate 
of ionizing photons, $Q_0 (\rm s^{-1})$, needed to explain the 
observed emission.  An O7 star with the $\sim 0.2\ Z_{\odot}$
metallicity of NGC~5253 \citep{b:walsh_mnras89} has $\log(Q_0) = 49.0$ 
\citep{b:smith_mnras02}.  This implies that radiation equivalent 
to that produced by $\gtrsim 160\ \pm 80 $\ O7 stars is needed to ionize the gas.  
This is a lower limit for the radiation escaping the nuclear starburst, 
since the flux will be affected by internal extinction and \fesc. 

NGC~5253 is host to many young clusters that could be 
the source for the ionizing radiation illuminating the 
cone.  Two likely candidates are NGC~5253-5 
\citep{b:calzetti_aj97} and the ionizing cluster of the 
radio supernebula \citep{b:beck_apj96, b:turner_apjl00, b:gorjian_apjl01}.  
The location of these objects, separated by just a few arcseconds 
on the sky \citep{b:alonso-herrero_apj04}, is noted by the cross in 
Figure \ref{f:ion_param}.  Both clusters are very young and very 
massive \citep{b:turner_apjl00, b:calzetti_aj97, b:alonso-herrero_apj04}. 
In fact, the radio supernebula may be gravitationally 
confined \citep{b:turner_nature03}.  The 20 pc region around the 
radio supernebula hosts an ionizing population $\sim$\ 7000 O7 
stars, with some uncertainty due to the high extinction 
towards this cluster \citep{b:turner_apjl04}.  Thus, the clusters in 
NGC~5253 easily generate enough ionizing flux to explain the ionization cone.  

\subsection{Kinematics} 

\begin{figure}
\centering
\includegraphics[width=8.5cm]{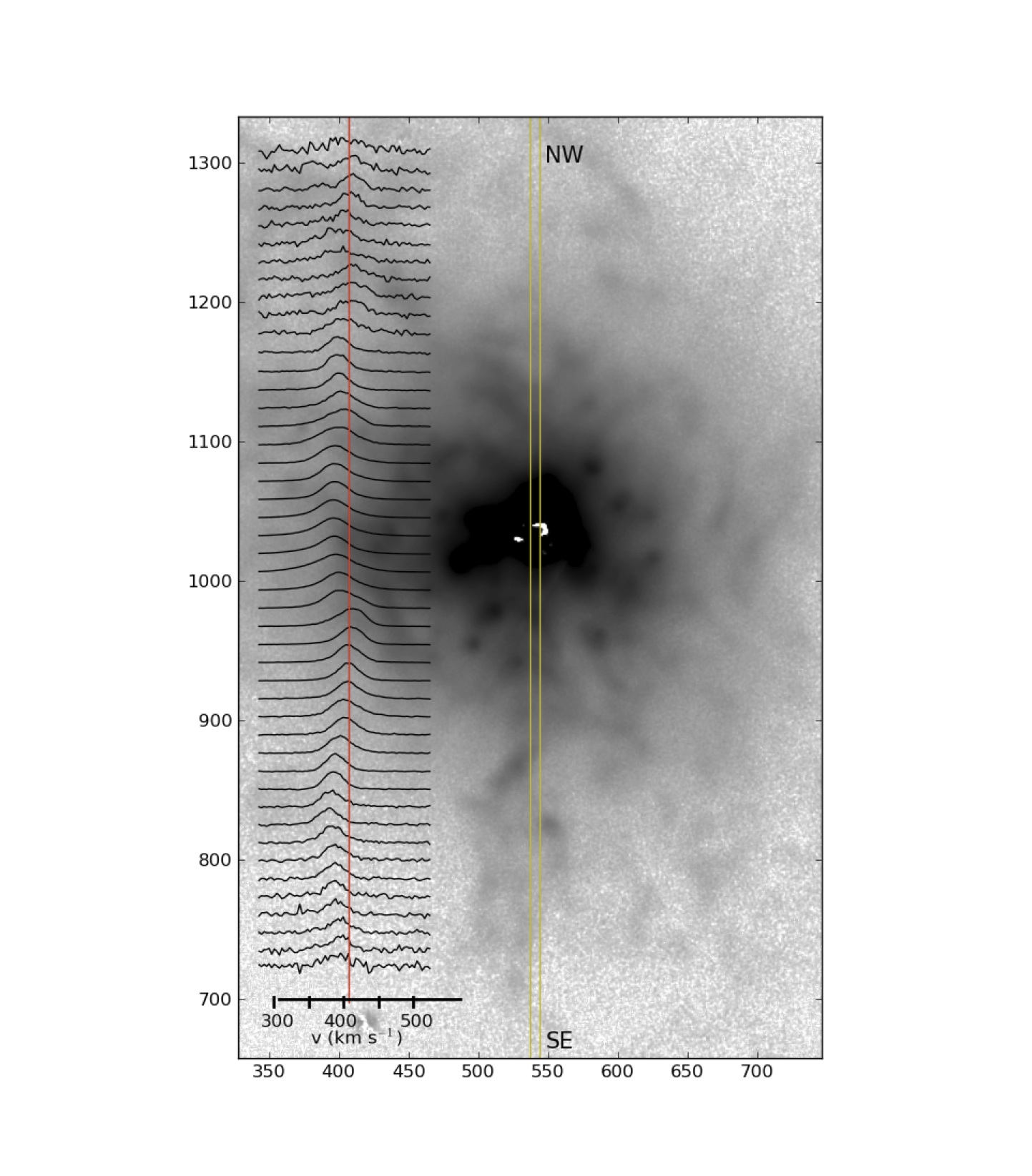}
\caption{NGC~5253 \Halpha\ image, minor axis oriented 
along the \emph{y}-axis.  \Halpha\ echelle profiles are 
overlaid to the left. The profiles are normalized to a peak of
one and the spacing between each profile is 2.\as6.  The red 
line denotes the $v_{sys}$ for NGC 5253, reported to be 
$407 \pm 3$\ \kms \citep{b:koribalski_aj04}.  
The slit position is marked in yellow. \label{f:kinematics}}
\end{figure}

The HI kinematics in NGC~5253 show only small scale rotation 
about the minor axis \citep{b:kobulnicky_aj08}, making it 
unlikely that NGC~5253 is rotationally supported \citep{b:caldwell_apj89}.  
In Figure \ref{f:kinematics}, the line profiles from an \Halpha\ 
echellegram \citep[Slit 3,][]{b:martin_apj95} are overlaid on 
the \Halpha\ image of NGC~5253.  The velocity shows 
significant variation with position, and a maximum 
projected velocity difference $\Delta v \sim 30\ \kms$ relative to systemic.  

An expanding wind might explain the kinematics.  In this 
scenario, we expect the velocity to increase with 
distance from the galaxy, since the wind is expanding 
into the lower density gas of the halo.  To the southeast, 
we do see an increase in velocity with galactocentric distance.
However, it reaches a maximum at $\sim$\ 40\as\ from the 
galaxy center before decreasing towards systemic at larger $R$.  
This does not strictly follow what is expected for a free-flowing 
wind, in which the velocity should continue to increase.  Additionally, 
an expanding wind typically exhibits line splitting, which is 
not seen in the \Halpha\ profiles \citep{b:martin_apj95}.  
Kinematic data from HI observations are suggestive 
of either inflow or outflow along the minor axis \citep{b:kobulnicky_aj08}.  
Therefore, although not completely straightforward, mechanical 
feedback might plausibly explain the observed kinematics based 
on the limited information available.

\subsection{Possible AGN}

Typically, ionization cones are associated with AGN, 
which suggests an alternative explanation for the
ionization cone in NGC~5253.  The morphology of the 
cone is very narrow.  This suggests collimation 
of the ionizing radiation, as would be expected if there is an AGN present.
The radio supernebula is particularly intriguing in this context.  
Only $\sim 0.\as6$ in size, it provides roughly $80\%$ of the 
galaxy's IR flux at 12\mum\ \citep{b:gorjian_apjl01}.  
There is some debate about whether a young super star cluster 
or AGN is the ionizing source of this nebula.  
Its radio continuum has a flat slope, consistent with 
thermal emission \citep{b:beck_apj96, b:turner_aj98, b:turner_apjl00},
but which does not rule out an AGN component \citep{b:turner_aj98}.
However, the observed spectral lines are also narrower than would 
be expected from an AGN \citep{b:beck_apj96}.  Additionally, 
X-ray observations of NGC 5253 show that the point sources and 
diffuse emission are consistent with star formation and shocks
\citep{b:ott_mnras05,b:summers_mnras04,b:martin_apj95}.  
Therefore, while we cannot rule out an AGN in NGC~5253, there 
are challenges facing that interpretation.

If NGC~5253 has an obscured AGN, a precessing jet might explain 
the gas kinematics.  In this case, the velocities on one side of 
the galaxy should be mirrored exactly on the opposite side.  In NGC~5253, 
the overall sinusoid in velocity to the southeast is roughly reflected 
to the northwest.  Along the ionization cone, the velocity 
transitions from redshift to blueshift around $\sim$\ 30\as\ from the 
galaxy.  To the northwest, the opposite transition, of roughly equal amplitude, 
is seen at $\sim$\ 28\as.  However, the reflection is not perfect; 
there are points to the northwest where the velocity briefly flips, which 
are not seen to the southeast. Additionally, the amount of blueshift to 
the northwest does not match the redshift observed in the southeast.  
Therefore, while the general kinematics could be explained by a 
precessing jet, the discrepancies are difficult to reconcile in detail.

\section{Implications: Orientation Bias}\label{s:impl}

Based on our detection of an 
ionization cone in NGC~5253, we suggest that an orientation bias 
can explain the difficulty in detecting escaping 
ionizing radiation in starburst galaxies.
We observe evidence that ionizing radiation from NGC 5253 is 
traveling to large radii, if not escaping the galaxy entirely.  Assuming 
the cone is axisymmetric, the estimated solid angle subtended by 
the cone is $\sim3\%$ of $4\pi$\ steradians, suggesting that the 
escape process happens over a small solid angle.  If we assume 
isotropic radiation from 7000 O7 stars, as above, we could expect to 
see ionized-gas emission in this solid angle equivalent to that 
produced by 210 O7 stars.  This is similar to our estimated $160\pm80$ O7 
stars needed to generate the observed \Halpha\ emission in the cone, 
supporting the physical link between the source and the cone.  
The narrow morphology suggests that in order to detect Lyman 
continuum, the line-of-sight to the galaxy must be close to the 
axis of escape.  Thus, the orientation of the 
galaxy should strongly influence the detectability of Lyman continuum
radiation. 

We can find further evidence for this scenario in recent 
observations by \citet{b:heckman_apj11}.  In their study of 
11 Lyman-break analog galaxies and 15 local starbursts, 
including NGC~5253, they found indirect evidence for significant 
\fesc\ in three of the Lyman-break analogs.  Outflow velocities 
on the order of $10^3 \rm\ \kms$ were found 
in all the galaxies suggested to have significant \fesc.  These outflow 
velocities are so high, that we can interpret it to mean the direction 
of the outflow must be closely aligned to the lines-of-sight towards 
the galaxies.  Since winds in starburst galaxies are generally 
coincident with the poles \citep{b:veilleux_araa05}, this suggests 
that these inferred high \fesc\ are all associated with face-on systems.  
This fits our explanation that measurements of large \fesc\ are biased towards 
galaxies whose outflow happens along our line-of-sight.

\section{Conclusion}

The ionization cone in NGC~5253 provides a new perspective 
to our understanding of the fate of ionizing photons in 
starburst galaxies.  This ionization cone appears optically thin, 
which is suggestive of the escape of ionizing radiation 
along the cone.  At minimum, ionizing radiation is escaping 
from the nuclear starburst into the galaxy's halo.  We considered both 
stellar and non-stellar sources for the ionization cone.  
Assuming a stellar source, the massive clusters in NGC~5253 
produce enough ionizing radiation to explain the \Halpha\ 
emission observed. However, the possibility of an obscured AGN 
has not been ruled out.  The \Halpha\ kinematics 
along the cone exhibit a complex morphology, but are consistent with 
some form of outflow. The data are not straightforward 
to interpret, but might be explained by an expanding wind 
or perhaps a precessing jet. 

Using the ionization cone of NGC~5253 as an analog for 
other starburst galaxies, we see that ionizing radiation escapes 
along the minor axis.  Additionally, the solid angle over which 
this radiation escapes is small, based on the cone morphology.  This will 
constrain the viewing angles from which Lyman continuum radiation can be detected. 
Thus, the orientation of the galaxy will strongly influence 
the ability to detect Lyman continuum radiation.  NGC~5253 may be the nearest 
starburst galaxy with a significant \fesc, and offers a unique 
opportunity to study the emission mechanisms in detail.

\acknowledgments
We kindly thank David Osip for recovering important data 
from the Magellan Archives, John Glaspey for converting 
old image formats to current ones, and Eric Pellegrini for useful discussions.  
We thank the referee for thorough and helpful 
comments.  This work was funded by NSF grants AST-0806476 to MSO and 
AST-1009583 to SV.

{\it Facilities:} \facility{Magellan:Baade()}


\clearpage

\end{document}